\documentclass[aps,nofootinbib,twocolumn,superscriptaddress]{revtex4-1}

\usepackage{epsfig}
\usepackage{color}
\usepackage[latin1]{inputenc}
\usepackage{float,amsmath}
\usepackage{graphicx}

\newcommand{\be}{\begin{eqnarray}}
\newcommand{\ee}{\end{eqnarray}}

\newcommand{\bfJ}{{\bf J}}

\newcommand{\benum}{\begin{enumerate}}
\newcommand{\eenum}{\end{enumerate}}

\newcommand{\calP}{{\cal P}}

\begin{document}

\title{Hydrodynamic evolution and jet energy loss in Cu+Cu collisions}

\author{Bj\"orn Schenke}
\affiliation{Physics Department, Bldg. 510A, Brookhaven National Laboratory, Upton, NY 11973, USA}
\affiliation{Department of Physics, McGill University, 3600 University Street, Montreal, Quebec, H3A\,2T8, Canada}

\author{Sangyong Jeon}
\affiliation{Department of Physics, McGill University, 3600 University Street, Montreal, Quebec, H3A\,2T8, Canada}

\author{Charles Gale}
\affiliation{Department of Physics, McGill University, 3600 University Street, Montreal, Quebec, H3A\,2T8, Canada}

\begin{abstract}
We present results from a hybrid description of Cu+Cu collisions using 3+1 dimensional hydrodynamics (\textsc{music}) for the bulk evolution and
a Monte-Carlo simulation (\textsc{martini}) for the evolution of high momentum partons in the hydrodynamical background.
We explore the limits of this description by going to small system sizes and determine the dependence on different fractions
of wounded nucleon and binary collisions scaling of the initial energy density.
We find that Cu+Cu collisions are well described by the hybrid description at least up to 20\% central collisions.  
\end{abstract}

\maketitle

\hyphenation{music martini}

\section{Introduction}
\label{intro}

In order to learn about the range of applicability of both hydrodynamics and energy loss mechanisms
in heavy-ion collisions, we study Cu+Cu collisions at $\sqrt{s}=200\,{\rm GeV}$ and different centralities.
Cu+Cu collisions are ideally suited for this because they are systems with a number of participants
that corresponds to that in peripheral Au+Au collisions \cite{Adler:2003qi}, and 
can be studied experimentally with reduced uncertainties \cite{Adare:2008cx}. One should thus be able to study the transition from ``soft'' (hydro-like) features to ``hard'' features (jet-like).

Hydrodynamics assumes that the system is equilibrated within a short time of the order of $\tau_0=0.5\,{\rm fm}/c$.
This assumption could be fulfilled for central Cu+Cu collisions but is unlikely to hold for peripheral ones.
Hence, it is important to study a wide range of centrality classes to see when and for which $p_T$ range the assumption breaks down.

In this work we are able to describe the entire experimentally covered $p_T$ range. 
The soft or low-$p_T$ sector of the matter produced is described using 3+1D relativistic hydrodynamics 
(\textsc{music} \cite{Schenke:2010nt}) 
and the hard sector ($p_T\gtrsim 2\,{\rm GeV}$) with the Monte-Carlo simulation \textsc{martini} \cite{Schenke:2009gb}, which
uses the hydrodynamic calculation as input and evolves the hard partons in the soft hydrodynamic background 
with the  McGill-AMY (Arnold, Moore and Yaffe) energy loss approach which treats radiative 
\cite{Arnold:2001ms,Arnold:2002ja}  and elastic processes \cite{Schenke:2009ik}. 

We can therefore achieve a complete description of Cu+Cu collisions under the assumption that both hydrodynamics and the energy loss formalism are valid. Knowing that we reach the limits of applicability of both models particularly for large centralities, 
this study allows to pin down observables that are sensitive to a failure of our approach and draw conclusions about the nature of
the produced system in Cu+Cu collisions.

\vspace{-0.25cm}
\section{Soft part - hydrodynamics}
\label{hydro}
To describe the soft part of the matter produced in a heavy-ion collision we employ hydrodynamics.
More precisely, we use the ideal 3+1D relativistic implementation of the Kurganov-Tadmor scheme \cite{Kurganov:2000} -- \textsc{music} -- introduced in \cite{Schenke:2010nt}.

The hydrodynamic equations have the following general form:
\be
\partial_t \rho_a = -\nabla{\cdot}\bfJ_a\,,
\label{eq:conserv_a}
\ee
where $a$ runs from 0 to 4, labelling the energy, 3 components of the momentum and the net baryon
density. 

The task is to solve these equations together with the equation of state. The Kurganov-Tadmor (KT) method, combined with a suitable flux limiter,  
is a non-oscillatory and simple central difference scheme 
with a small artificial viscosity that can also handle shocks very well
It is Riemann-solver free and hence does not require calculating the local characteristics. 
 
 When describing heavy-ion collisions it is useful to employ $\tau-\eta_s$ coordinates
 defined by 
 \begin{eqnarray}
   t &=& \tau\cosh\eta_s\,,\nonumber\\
   z &=& \tau\sinh\eta_s\,.
   \label{eq:tz}
 \end{eqnarray}
 Then the conservation equation
 $\partial_\mu J^\mu = 0$ becomes
 \be
  \partial_\tau (\tau J^\tau)
 +
 \partial_{\eta_s} J^{\eta_s}
 + \partial_v(\tau J^v)& = &0\,,
 \label{eq:jcons}
 \ee
 where
 \be
 J^\tau &= & (\cosh\eta_s J^0 - \sinh\eta_s J^3)\,,
 \label{eq:jtau}
 \\
 J^{\eta_s} &= &
 ( \cosh\eta_s J^3 - \sinh\eta_s J^0 )\,,
 \label{eq:jeta}
 \ee
 which is nothing but a Lorentz boost with the space-time rapidity
 $\eta_s = \tanh^{-1}(z/t)$. 
 The indices $v$ and $w$ refer to the transverse $x,y$ coordinates which are not
 affected by the boost.
 Applying the same transformation to both indices of $T^{\mu\nu}$,
 one obtains
\be
\partial_\tau (\tau T^{\tau\tau})
+ \partial_{\eta_s} (T^{\eta_s\tau}) + \partial_v (\tau T^{v\tau})
+ T^{\eta_s\eta_s} = 0\,,
\label{eq:econs}
\ee
and
\be
\partial_\tau (\tau T^{\tau\eta_s})
+ \partial_{\eta_s}(T^{\eta_s\eta_s})
+ \partial_v(\tau T^{v\eta_s}) + T^{\tau\eta_s} = 0\,,
\label{eq:etacons}
\ee
and
\be
\partial_\tau (\tau T^{\tau v})
+ 
\partial_{\eta_s} (T^{\eta_s v})
+
\partial_w (\tau T^{w v})
& = &
0\,,
\label{eq:xycons}
\ee
These 5 equations, namely Eq.\,(\ref{eq:jcons}) for the net baryon current, and
Eqs.\,(\ref{eq:econs}, \ref{eq:etacons}, \ref{eq:xycons}) for the energy and
momentum are the equations we solve with the KT scheme
described in detail in \cite{Kurganov:2000,Schenke:2010nt}.


To close the set of equations (\ref{eq:jcons}, \ref{eq:econs}, \ref{eq:etacons}, \ref{eq:xycons}) we must provide a nuclear equation of 
state $\calP(\varepsilon,\rho)$
which relates the local thermodynamic quantities.
The presented calculations all employ an equation of state extracted from recent lattice QCD calculations \cite{Huovinen:2009yb}.
There, several parametrizations of the
equation of state which interpolate between the lattice data at high temperature and a hadron-resonance gas
in the low temperature region were constructed. 
We adopt the parametrization ``s95p-v1'' (and call it EOS-L in the following), where
the fit to the lattice data was done above $T = 250\,{\rm  MeV}$, 
and the entropy density was constrained at $T = 800\,{\rm  MeV}$ to be 
95\% of the Stefan-Boltzmann value. Furthermore, one "data-point" was added to the fit to make the peak in the trace anomaly higher. 
See Ref. \cite{Huovinen:2009yb} for more details on this parametrization of the nuclear equation of state.


The initialization of the energy density is done using the Glauber model (see Ref. \cite{Miller:2007ri} and references therein):
Before the collision the density distribution of the two nuclei is described by
a Woods-Saxon parametrization
\begin{equation}
  \rho_A(r) = \frac{\rho_0}{1+\exp[(r-R)/d]}\,,
\end{equation}
with $R=4.163\,{\rm fm}$ and $d=0.606\,{\rm fm}$ for Cu nuclei.
The normalization factor $\rho_0$ is set such that  $\int d^3r \rho_A(r)=A$. 
With the above parameters we get $\rho_0=0.17\,{\rm fm}^{-3}$.

The \emph{nuclear thickness function}
\begin{equation}\label{thickness}
  T_A(x,y) = \int_{-\infty}^{\infty}dz\,\rho_A(x,y,z)\,,
\end{equation}
where $r=\sqrt{x^2+y^2+z^2}$, 
can be used to express both the density of wounded nucleons $n_{\rm WN}$ and binary collisions $n_{\rm BC}$ (see e.g. Ref. \cite{Schenke:2010nt}).
Whether the deposited energy density or entropy density scales with $n_{\rm WN}$ or $n_{\rm BC}$ is not
clear from first principles. SPS data suggests that the final state particle multiplicity is proportional to the number of wounded 
nucleons. At RHIC energies a violation of this scaling was found (the particle production per wounded nucleon is a function increasing 
with centrality. This is attributed to a significant contribution from hard processes, scaling with the number of binary collisions).

As in \cite{Schenke:2010nt} the shape of the initial energy density distribution in the transverse plane is parametrized as
\begin{equation}\label{inieps}
  W(x,y,b)=(1-\alpha)\, n_{\rm WN}(x,y,b)+\alpha\, n_{\rm BC}(x,y,b)\,,
\end{equation}
where $\alpha$ determines the fraction of the contribution from binary collisions, and we will study the effect of different choices for $\alpha$.

For the longitudinal profile we employ the prescription used in Refs. 
\cite{Ishii:1992xi,Morita:1999vj,Hirano:2001yi,Hirano:2001eu,Hirano:2002ds,Morita:2002av,Nonaka:2006yn}.
It is composed of a flat region around $\eta_s=0$, and of half a Gaussian in the forward
and backward direction:
\begin{equation}
  H(\eta_s)=\exp\left[-\frac{(|\eta_s|-\eta_{\rm flat}/2)^2}{2 \sigma_{\eta}^2}\theta(|\eta_s|-\eta_{\rm flat}/2)\right]\,.
\end{equation}

The full energy density distribution is then given by
\begin{equation}
  \varepsilon(x,y,\eta_s,b) = \varepsilon_0\, \, H(\eta_s)\,W(x,y,b)/W(0,0,0)\,.
\end{equation} 

The parameters $\eta_{\rm flat}$ and $\sigma_\eta$ are tuned to data and will be quoted below.

The impact parameter $b$ is taken to be the average impact parameter $\langle b \rangle$ for a given centrality class and determined using
the optical Glauber model (see e.g. \cite{Niemi:2008ta}). The obtained values are listed in Table \ref{table:b}.

\begin{table}[tb]
\centering
\begin{tabular}{|c|c|}	
\hline
centrality \%	& $\langle b\rangle$ [fm]\\ 
\hline
0-5	& 1.61	\\
0-6     & 1.76  \\
0-10    & 2.28  \\
5-10	& 2.95	\\
6-15    & 3.47  \\
10-15	& 3.83	\\
10-20   & 4.16  \\
15-20	& 4.55	\\
15-25   & 4.82  \\
20-30	& 5.4	\\
25-35   & 5.92  \\
30-40	& 6.4	\\
\hline
\end{tabular}
\caption{Average impact parameters in selected centrality classes from the optical Glauber model.\label{table:b}} 
\end{table}

The spectrum of produced hadrons of species $i$ with degeneracy $g_i$ is given by the Cooper-Frye formula \cite{Cooper:1974mv}:
\begin{equation}\label{cf}
E\frac{dN}{d^3p}=\frac{dN}{dy p_T dp_T d\phi_p} = g_i \int_\Sigma f(u^\mu p_\mu) p^\mu d^3\Sigma_\mu\,,
\end{equation}
with the distribution function
\begin{equation}
  f(u^\mu p_\mu) = \frac{1}{(2\pi)^3}\frac{1}{\exp((u^\mu p_\mu -\mu_i)/T_{\rm FO})\pm 1}\,,
\end{equation}
where $T_{\rm FO}$ is the freeze-out temperature, $u^\mu$ is the flow velocity of the fluid,
and the energy is given by the invariant expression $E=u^\mu p_\mu$.
In practice, we determine the freeze-out hyper-surface geometrically using a triangulation of the three dimensional hyper-surface that
is embedded in four-dimensional space-time. The details of this algorithm are described in Ref. \cite{Schenke:2010nt}.

\section{Hard part - Monte-Carlo jet evolution}
\label{mc}
The hard part of the system is described using the Monte-Carlo simulation \textsc{martini} \cite{Schenke:2009gb}.
At the core of \textsc{martini} lies the McGill-AMY formalism for jet evolution in a dynamical thermal medium.
This evolution is governed by a set of coupled Fokker-Planck-type rate equations of the form
\begin{eqnarray}\label{jet-evolution-eq}\hspace{-0.3cm}
\frac{dP(p)}{dt}\!=\!\int_{-\infty}^{\infty}\!\!\!\!\!\!dk
\left(\!P(p{+}k) \frac{d\Gamma(p{+}k,k)}{dk} - P(p)\frac{d\Gamma(p,k)}{dk}\!\right)\,,
\end{eqnarray}
where ${d\Gamma(p,k)}/{dk}$ is the transition rate for
processes where partons of energy $p$ lose energy $k$. 

In the finite temperature field theory approach of  AMY,
radiative transition rates can be calculated by means of integral equations
\cite{Arnold:2002ja} which correctly reproduce both the Bethe-Heitler and the LPM results
in their respective limits \cite{Jeon:2003gi}. 
Transition rates for the processes $g\rightarrow gg$, $q(\bar{q})\rightarrow q(\bar{q}) g$,
and $g\rightarrow q\bar{q}$ are included as well as elastic processes employing the transition rates computed in
\cite{Schenke:2009ik}. Furthermore, gluon-quark and quark-gluon conversion
due to Compton and annihilation processes, as well as the QED processes of
photon radiation $q\rightarrow q\gamma$ and jet-photon conversions \cite{Fries:2002kt,Turbide:2007mi} are implemented.

In \textsc{martini}, Eq.\,(\ref{jet-evolution-eq}) is solved using Monte Carlo methods, keeping track 
of each individual parton, rather than the probability distributions $P$.
This method provides information on the full microscopic event configuration in the high momentum regime, including correlations,
which allows for a very detailed analysis and offers a direct interface between theory and experiment.

In a full heavy-ion event, the number of individual nucleon-nucleon collisions that produce partons with a certain minimal
transverse momentum $p_T^{\rm min}$ is determined from the total inelastic cross-section, provided by \textsc{pythia}, which
is also used to generate those individual hard collisions. 
The initial transverse positions of the hard collisions
are determined by the initial jet density distribution 
\begin{eqnarray}
\mathcal{P}_{AB}(b,\mathbf{r}_\perp) &=& \frac{T_A(\mathbf{r}_\perp + \mathbf{b}/2)T_B(\mathbf{r}_\perp -
\mathbf{b}/2)}{T_{AB}(b)}\,,
\end{eqnarray}
which is determined by the nuclear thickness (\ref{thickness}) and overlap functions
\begin{eqnarray}
T_{AB}(b) =\int d^2r_\perp T_A(\mathbf{r}_\perp) T_B(\mathbf{r}_\perp+\mathbf{b})\,.
\end{eqnarray}
The initial parton distribution functions can be selected with the help of the Les Houches Accord PDF Interface (LHAPDF) 
\cite{Whalley:2005nh}.
Isospin symmetry is assumed and nuclear effects on the parton distribution functions are included
using the EKS98 \cite{Eskola:1998df} parametrization.

As discussed above, the soft medium is described by hydrodynamics.
Before the hydrodynamic evolution begins ($\tau<\tau_0$), the partons shower as in vacuum.
During the medium evolution, individual partons move through the background according to their velocity.
Probabilities to undergo an interaction are determined in the local fluid cell rest frame using the transition rates and 
the local temperature. The boost into the local fluid rest-frame is performed using the flow velocities provided by the hydro calculation.

If a process occurs, the radiated or transferred energy is sampled from the transition rate of that process.
In case of an elastic process, the transferred transverse momentum is also sampled, while
for radiative processes collinear emission is assumed.

Radiated partons are further evolved if their momentum is above a certain threshold $p_{\rm min}\simeq 2-3\,{\rm GeV}$. 
The overall evolution of a parton stops once its energy in a fluid cell's rest frame falls below the limit
of $4T$, where $T$ is the local temperature.
For partons that stay above that threshold, the evolution ends once they enter the hadronic phase of the background medium.
In the mixed phase or a crossover region, processes occur only for the QGP fraction.
When all partons have left the QGP phase, hadronization is performed by \textsc{pythia}, to which the complete
information on all final partons is passed. Because \textsc{pythia} uses the Lund string fragmentation model 
\cite{Andersson:1983ia,Sjostrand:1984ic}, it is essential to
keep track of all color strings during the in-medium evolution. 
For more information on the details of \textsc{martini} please refer to \cite{Schenke:2009gb}.

\section{Results}
\label{results}
We begin by presenting results for the soft transverse momentum region obtained entirely by \textsc{music}.
We always compare three different parametrizations, which essentially differ by the contribution of binary collision scaling of the initial
energy density distribution (see Eq. (\ref{inieps})). The details of the parameter sets are listed in Table \ref{tab:parms}.

\begin{center}
\begin{table*}
  \begin{tabular}{|c|c|c|c|c|c|c|}
    \hline
    $\alpha$ &  $\tau_0 [{\rm fm}]$ & $\varepsilon_0 [{\rm GeV}/{\rm fm}^3]$ &      $\varepsilon_{\rm FO} [{\rm GeV}/{\rm fm}^3]$ & $T_{\rm FO} [{\rm MeV}]$  
    & $\eta_{\rm flat}$ & $\sigma_{\eta}$ \\\hline
     0.1      &0.55                & 20                                                   & 0.14 & $\approx 140$
    & 4      & 1.2 \\
     0.5      &0.55                & 26                                                   & 0.14 & $\approx 140$
    & 4      & 1.2 \\
     1        &0.55                & 29                                                   & 0.14 & $\approx 140$
    & 4      & 1.2 \\
    \hline
  \end{tabular}
  \caption{Parameter sets. \label{tab:parms}}
\end{table*}
\end{center}

\begin{figure}[tb]
  \begin{center}
    \includegraphics[width=8.5cm]{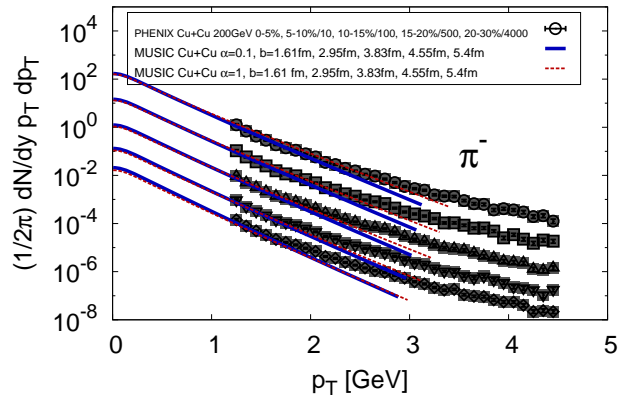}
    \caption{Transverse momentum spectra of negative pions for different centrality classes compared to preliminary PHENIX data. The collisions go from central (top) to more peripheral (bottom).
      \label{fig:pt-CuCu}}
  \end{center}
\end{figure}

\begin{figure}[tb]
  \begin{center}
    \includegraphics[width=8.5cm]{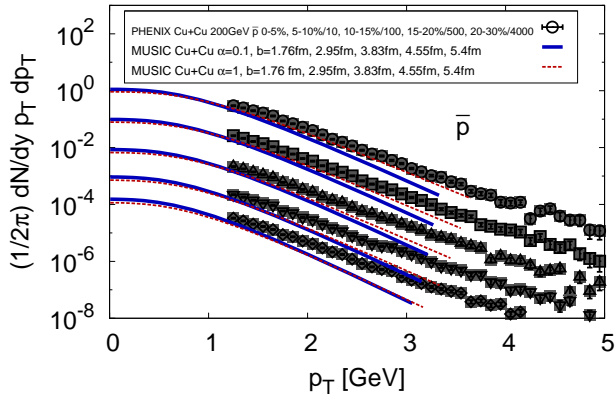}
    \caption{Transverse momentum spectra of anti-protons for different centrality classes compared to preliminary PHENIX data.  
      \label{fig:proton-pt-CuCu}
      The experimental data is underestimated because of the assumption of chemical equilibrium.}
  \end{center}
\end{figure}

Fig. \ref{fig:pt-CuCu} shows the transverse momentum spectra of negative pions for different centralities compared to preliminary data from PHENIX (see e.g. \cite{Chujo:2007hi}).
For the most central collisions both parametrizations describe the experimental data reasonably well for $p_T\lesssim 2\,{\rm GeV}$.
The larger the impact parameter, the lower the $p_T$ at which the description using hydrodynamics fails.
For 20-30\% central collisions the calculation begins to deviate from the experimental data at approximately $p_T\gtrsim 1.5\,{\rm GeV}$.
Generally, the calculation with larger $\alpha$ leads to harder spectra. 
For anti-protons we find a very similar behavior, shown in Fig. \ref{fig:proton-pt-CuCu}. In this case the result is generally below the data
because of the assumed chemical equilibrium (see Ref. \cite{Schenke:2010nt}), the overall shape of the spectra is however
well reproduced for low $p_T$.

\begin{figure}[tb]
  \begin{center}
    \includegraphics[width=8.5cm]{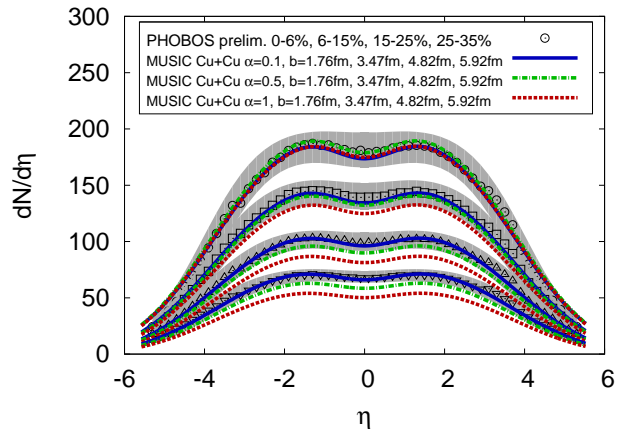}
    \caption{Centrality dependence of pseudorapidity distributions of charged hadrons compared to preliminary PHOBOS data \cite{Veres:2008nq} (open symbols).
      Calculations using $\alpha=0.1$ describe all centrality classes very well, while larger $\alpha$ lead to deviations for larger centralities. 
      \label{fig:etaCuCu}}
  \end{center}
\end{figure}

Although we find harder spectra in the case of $\alpha=1$ it is unclear which parametrization is favored solely by studying the
$p_T$ spectra of pions and anti-protons. The centrality dependence of charged particle pseudo-rapidity distributions in Fig. \ref{fig:etaCuCu} reveals 
a much more obvious difference. While in the case $\alpha=0.1$ the experimental data is very well described for all shown centrality classes,
the calculations using $\alpha=0.5$ and $\alpha=1$ only describe the most central data to which the parameters were adjusted, 
but differ increasingly with increasing impact parameter.

It is remarkable how well the $p_T$ integrated pseudorapidity distributions are described even for very large pseudorapidities and impact parameters.
This indicates that for low $p_T$, i.e., $p_T\lesssim 1\,{\rm GeV}$, hydrodynamics works well, hinting at that the low momentum modes 
equilibrate fast enough even in small systems, while higher momentum modes do not.

\begin{figure}[tb]
  \begin{center}
    \includegraphics[width=8.5cm]{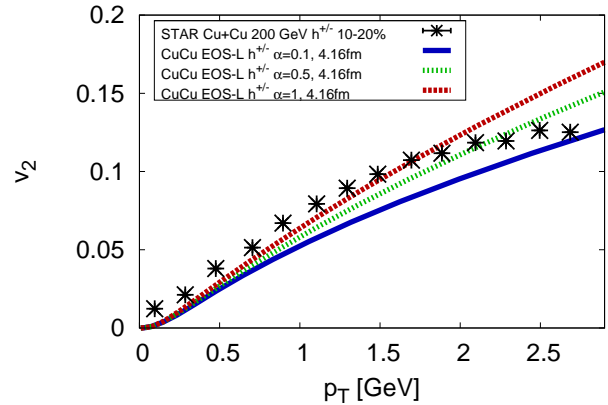}
    \caption{Elliptic flow coefficient $v_2$ as a function of transverse momentum for 10-20\% central events compared to STAR data \cite{Abelev:2010tr}. \label{fig:v2CuCu}}
  \end{center}
\end{figure}

Next, we present the elliptic flow parameter $v_2$ and compare to STAR data \cite{Abelev:2010tr}.
Because we do not include fluctuations, we do not expect the centrality dependence of elliptic flow to be reproduced. 
See Ref. \cite{Schenke:2010rr} for a demonstration of the importance of fluctuations on elliptic flow observables.
In Fig. \ref{fig:v2CuCu} we show results for 10-20\% central events, a centrality class in which $v_2(p_T)$ was found to be well described 
by smooth initial conditions, at least for Au+Au collisions \cite{Schenke:2010rr}.
The dependence on the $\alpha$ parameter is as expected. Larger $\alpha$ meaning higher initial eccentricity, leads to stronger elliptic flow.
To draw further conclusions from this comparison a detailed event-by-event study is needed.

Extracting the evolution information from the hydrodynamic simulation, i.e., the temperature and flow velocities as functions of
time and spatial coordinates, we can use this information as an input for the simulation of the evolution of the hard degrees of freedom,
\textsc{martini}. 
 
\begin{figure}[tb]
  \begin{center}
    \includegraphics[width=8.5cm]{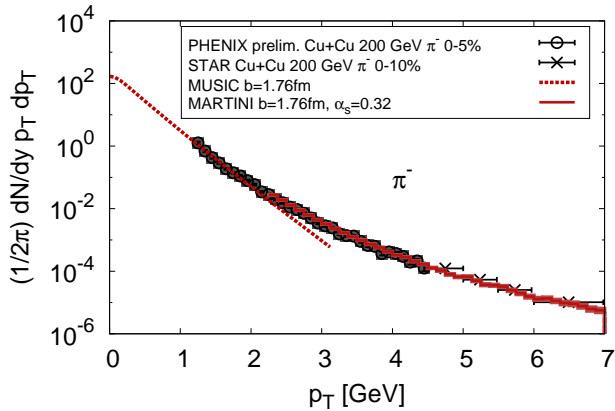}
    \caption{Transverse momentum spectrum of negative pions from the hydrodynamic calculation \textsc{music} (dashed line) 
      and \textsc{martini} using \textsc{music} input (solid line) compared to data from PHENIX (preliminary) 
      and STAR \cite{Abelev:2009ue}. 
      \label{fig:pt-MM-CuCu}}
  \end{center}
\end{figure}

This way we can compute the hard part of the spectrum and combine results for the soft and hard part as done in 
Fig.\,\ref{fig:pt-MM-CuCu} for negative pions. 
The additional free parameter that regulates the amount of energy loss in \textsc{martini} is the strong coupling constant,
which we fix at $\alpha_s=0.32$. The value of transverse momentum up to where hydrodynamics alone provides a good description and
beyond which the contribution from jets becomes important decreases with decreasing system size. For central Au+Au collisions
the value lies between $3$ and $4\,{\rm GeV}$ \cite{Schenke:2010nt}, while Fig.\,\ref{fig:pt-MM-CuCu} shows that for Cu+Cu it lies 
around $2\,{\rm GeV}$ for central
collisions, and from Fig.\,\ref{fig:pt-CuCu} one can read off that it decreases more when going to larger centralities.

Next, we study the nuclear modification factor for neutral pions, defined by
\begin{equation}
  R_{AA}=\frac{1}{N_{\rm coll}(b)} \frac{dN_{AA}(b)/d^2p_T dy}{dN_{pp}/d^2p_Tdy}\,,
\end{equation}
where $N_{\rm coll}(b)$ is the number of binary collisions at given impact parameter $b$.
The reference spectrum $dN_{pp}/d^2p_Tdy$ for p+p collisions was computed in \cite{Schenke:2009gb}.

\begin{figure}[tb]
  \begin{center}
    \includegraphics[width=8.2cm]{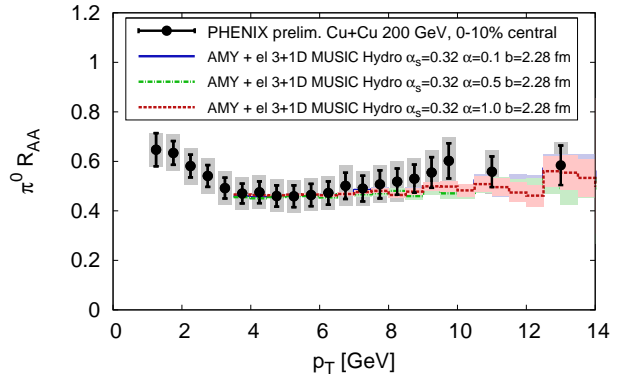}
    \caption{Nuclear modification factor for neutral pions in 0-10\% central Cu+Cu collisions compared to preliminary data from PHENIX.    
      \label{fig:raa0-10} }
  \end{center}
\end{figure}

\begin{figure}[tb]
  \begin{center}
    \includegraphics[width=8.5cm]{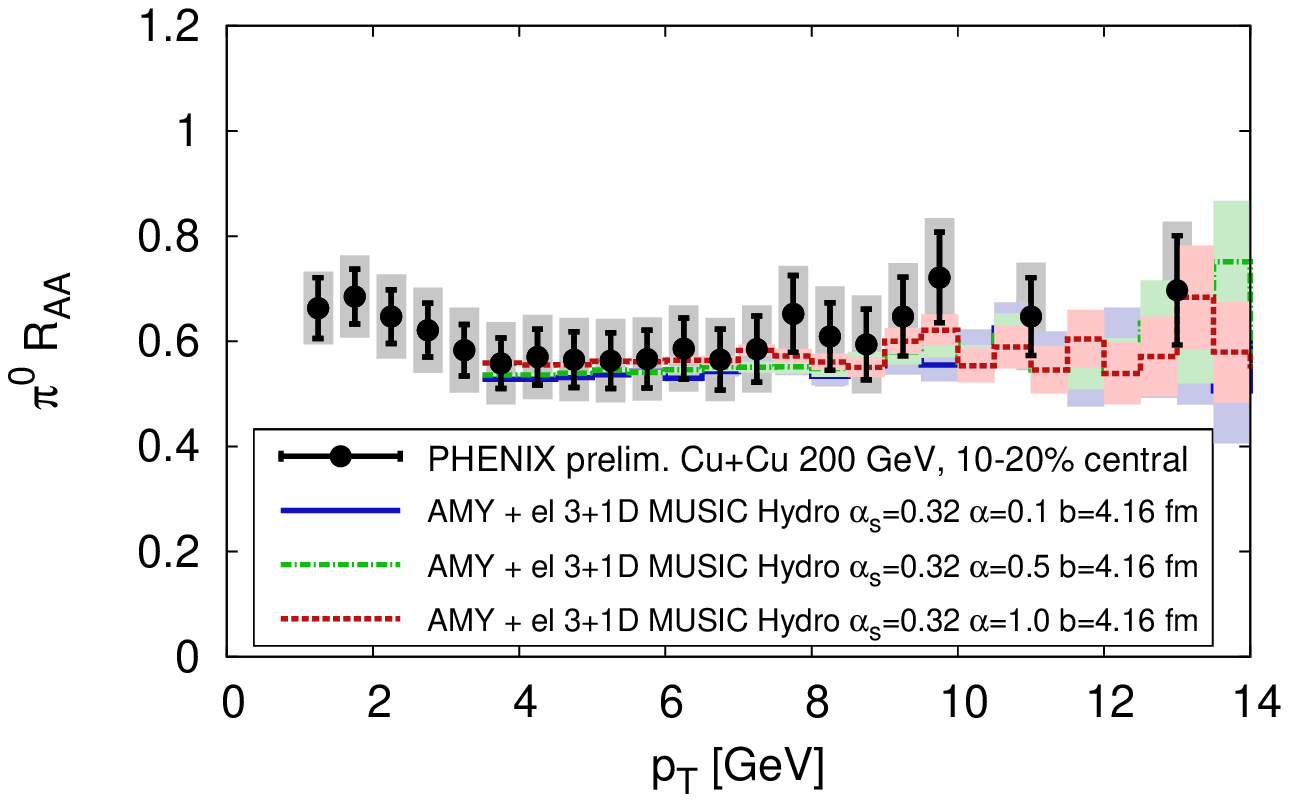}
    \caption{Nuclear modification factor for neutral pions in 10-20\% central Cu+Cu collisions compared to preliminary data from PHENIX.    
      \label{fig:raa10-20} }
  \end{center}
\end{figure}

\begin{figure}[tb]
  \begin{center}
    \includegraphics[width=8.5cm]{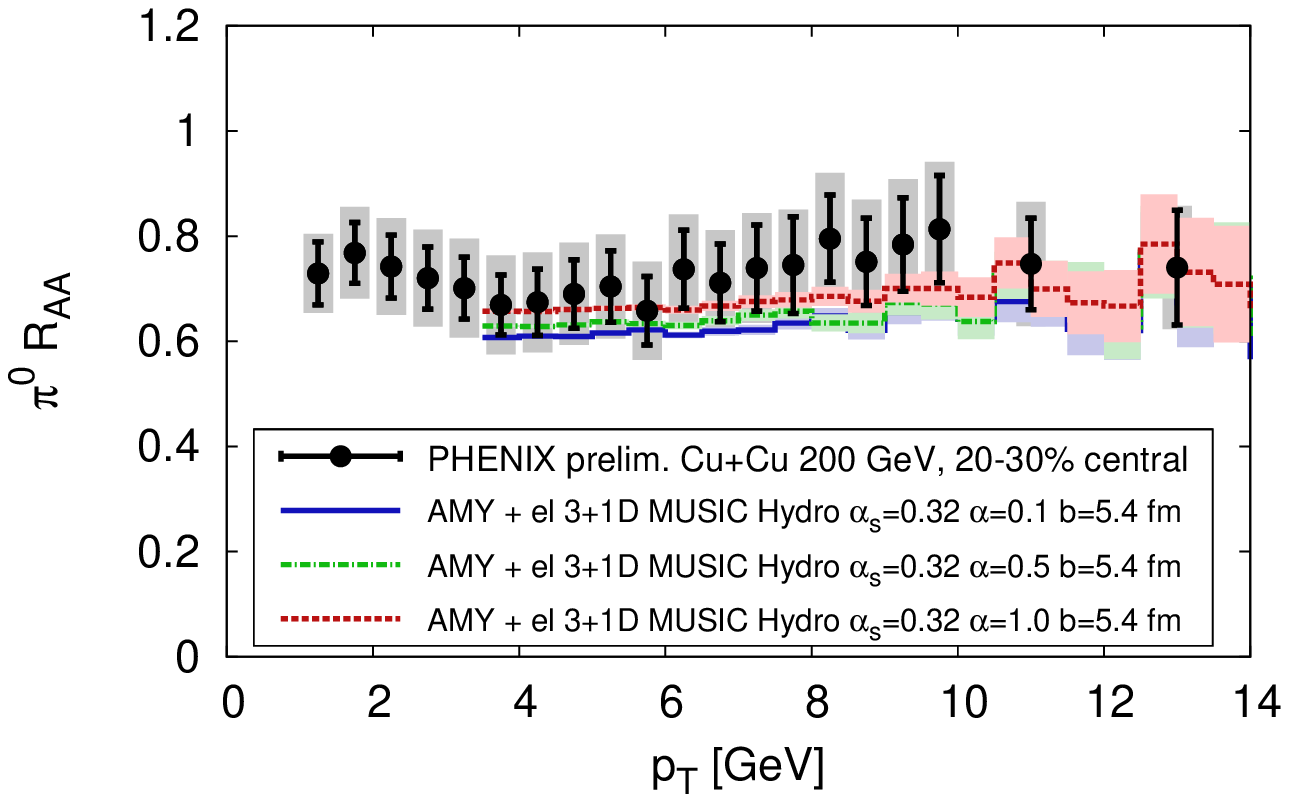}
    \caption{Nuclear modification factor for neutral pions in 20-30\% central Cu+Cu collisions compared to preliminary data from PHENIX.    
      \label{fig:raa20-30} }
  \end{center}
\end{figure}

\begin{figure}[tb]
  \begin{center}
    \includegraphics[width=8.5cm]{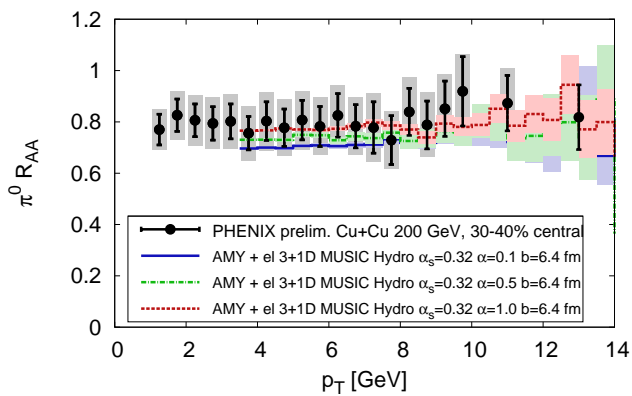}
    \caption{Nuclear modification factor for neutral pions in 30-40\% central Cu+Cu collisions compared to preliminary data from PHENIX.    
      \label{fig:raa30-40} }
  \end{center}
\end{figure}

Figures \ref{fig:raa0-10} to \ref{fig:raa30-40} show the nuclear modification factor $R_{AA}$ for different centrality classes. The value of 
$\alpha_s$ was chosen to match the experimental data in the most central collisions best, but when going to larger centrality classes
the impact parameter is the only quantity that is changed. We see that all hydrodynamic backgrounds ($\alpha\in\{0.1,0.5,1\}$) 
lead to good agreement with the experimental data in the two most central bins 
(Figs.\,\ref{fig:raa0-10} and \ref{fig:raa10-20}).
All calculations, but most significantly the one using $\alpha=0.1$, deviate from the 
20-30\% and 30-40\% central experimental data.
This indicates that the combination of ideal hydrodynamics with our energy loss mechanism 
begins to fail to describe Cu+Cu systems at centralities $\gtrsim 20\%$.
Remember that hydrodynamics using $\alpha=0.1$ described the bulk data well, while deviations from experimental data
at more peripheral collisions are large for $\alpha=0.5$ and $\alpha=1$ (Fig.\,\ref{fig:etaCuCu}).

On the other hand $\alpha=1$ works best for $R_{AA}$. This indicates that the energy loss mechanism itself leads to a 
too weak centrality dependence when using the ``correct'' hydrodynamic background ($\alpha=0.1$). 
Using a hydro-background with a larger centrality dependence
of the density ($\alpha=1$) then works against this problem. Note, however, that for all $\alpha$ the results for $R_{AA}$
lie within the experimental error bars.
The effect of non-zero viscosity and event-by-event fluctuations \cite{Schenke:2010rr} remains to be studied.

Finally we study high momentum photon production in Cu+Cu collisions.
Photons in the high $p_T$ region produced in nuclear collisions are dominantly
direct photons, fragmentation photons, and jet-plasma photons. 
Direct photons are included in \textsc{pythia}.
Apart from leading order direct photons, \textsc{pythia} produces additional photons emitted during the vacuum showers,
which in part overlap with photons from next-to-leading order calculations. 
Fragmentation photons are also part of those produced in the shower. 
For heavy-ion reactions \textsc{martini} adds the very relevant jet-medium photons from photon radiation 
$q\rightarrow q\gamma$ (bremsstrahlung photons) and jet-photon conversion via 
Compton and annihilation processes.
The full vacuum shower is included in the calculation - most of the shower photons will be emitted
before the medium has formed and the parton has realized that it has formed. 
For more details on the implementation see \cite{Schenke:2009gb}.

\begin{figure}[tb]
  \begin{center}
    \includegraphics[width=8.5cm]{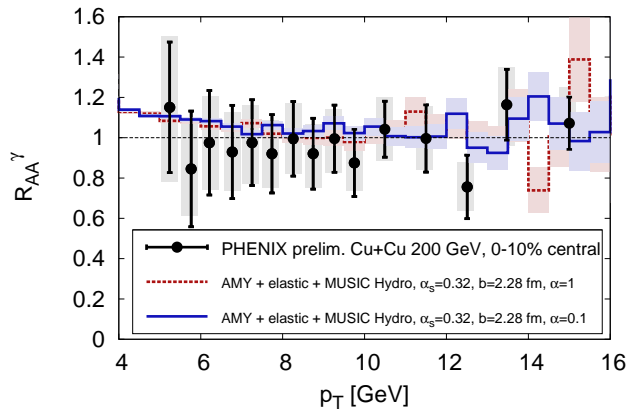}
    \caption{Nuclear modification factor for photons in 0-10\% central Cu+Cu collisions compared to preliminary data from PHENIX.
      \label{fig:photon-raa}}
  \end{center}
\end{figure}

In Fig.\,\ref{fig:photon-raa} we present 
the nuclear modification factor for photons compared to preliminary data from PHENIX (also see \cite{Chujo:2007hi}) in central collisions. 
Again, the reference spectrum from p+p collisions was computed in \cite{Schenke:2009gb}. 
The data is consistent with $R_{AA}^\gamma=1$ and so is the calculation for a wide range of $p_T$. 
The rise above one for $p_T\lesssim 6\,{\rm GeV}$ stems from the contribution of in-medium photon production.
Unfortunately the current experimental data does not allow to distinguish whether this contribution is present or not. 

\section{Conclusions and Outlook}
\label{conclusions}
We studied the applicability of both a 3+1D hydrodynamic model and the McGill-AMY jet energy loss and evolution scheme
to Cu+Cu collisions of different centralities. Because the created system is smaller than in Au+Au collisions
the limits of the models are already reached at a smaller impact parameter. The precise determination 
of these limits and the determination of observables that are sensitive to the failing of the models was the main 
objective of this work.

The hydrodynamic evolution was computed using the relativistic 3+1D implementation of the Kurganov-Tadmor scheme, dubbed \textsc{music},
while the evolution of the hard degrees of freedom was simulated using the Monte-Carlo simulation \textsc{martini}.
With these, we were able to compare calculations to experimental observables over the full range of measured transverse momentum.
 
As expected, the quality of the description of transverse momentum spectra of identified hadrons using only hydrodynamics
depends on the system size. While in Au+Au collisions the spectra in central collisions are typically 
well described up to $p_T\approx 3\,{\rm GeV}$, in Cu+Cu we find deviations at approximately $2\,{\rm GeV}$ for central collisions,
and lower $p_T$ for larger centralities. 

The hydrodynamic calculation of the centrality dependence of pseudorapidity distributions agrees very well with the experimental
data when we use 90\% wounded nucleon scaling and 10\% binary collisions scaling of the initial energy density distribution.
The parameter set using 100\% binary collision scaling does not reproduce the centrality classes above 10\%.
We therefore conclude that mostly wounded nucleon scaling provides a better description of experimental data. In this case,
the result shows that the low transverse momentum modes that dominate the distribution are thermalized, even in 30-40\% central collisions,
which is in agreement with the result for the transverse momentum spectra.

Including the contribution from jets by employing \textsc{martini}, which uses the hydrodynamic evolution as input, we were
able to describe the complete pion spectra in central collisions. The study of $R_{AA}$ for pions allows to pin down the
point where the combined ideal hydrodynamics+McGill-AMY evolution breaks down. 

The trend seen in the description of the centrality dependence of $R_{AA}$ agrees least with experimental data 
when using $\alpha=0.1$, which is preferred by the low momentum spectra.
While calculations for 0-10\% and 10-20\% central bins
agree well with experimental data, deviations begin for centralities larger than 20\%. 
This might indicate that the used energy loss
mechanism does not have the correct density and/or system size dependence for the small systems studied here.
The current error bars do not permit a more precise quantitative statement.

\section*{Acknowledgments}

B.P.S. thanks Roy Lacey for help with locating the experimental data.
This work was supported in part by the Natural Sciences and Engineering Research Council of Canada.
B.P.S.\ gratefully acknowledges a Richard H.~Tomlinson Postdoctoral Fellowship by McGill University.
B.P.S. was supported in part by the US Department of Energy under DOE Contract No.DE-AC02-98CH10886, and 
by a Lab Directed Research and De\-velopment Grant from Brookhaven Science Associates. 

\bibliography{hydro}

\begin{thebibliography}{31}%
\makeatletter
\providecommand \@ifxundefined [1]{%
 \@ifx{#1\undefined}
}%
\providecommand \@ifnum [1]{%
 \ifnum #1\expandafter \@firstoftwo
 \else \expandafter \@secondoftwo
 \fi
}%
\providecommand \@ifx [1]{%
 \ifx #1\expandafter \@firstoftwo
 \else \expandafter \@secondoftwo
 \fi
}%
\providecommand \natexlab [1]{#1}%
\providecommand \enquote  [1]{``#1''}%
\providecommand \bibnamefont  [1]{#1}%
\providecommand \bibfnamefont [1]{#1}%
\providecommand \citenamefont [1]{#1}%
\providecommand \href@noop [0]{\@secondoftwo}%
\providecommand \href [0]{\begingroup \@sanitize@url \@href}%
\providecommand \@href[1]{\@@startlink{#1}\@@href}%
\providecommand \@@href[1]{\endgroup#1\@@endlink}%
\providecommand \@sanitize@url [0]{\catcode `\\12\catcode `\$12\catcode
  `\&12\catcode `\#12\catcode `\^12\catcode `\_12\catcode `\%12\relax}%
\providecommand \@@startlink[1]{}%
\providecommand \@@endlink[0]{}%
\providecommand \url  [0]{\begingroup\@sanitize@url \@url }%
\providecommand \@url [1]{\endgroup\@href {#1}{\urlprefix }}%
\providecommand \urlprefix  [0]{URL }%
\providecommand \Eprint [0]{\href }%
\@ifxundefined \urlstyle {%
  \providecommand \doi  [0]{\begingroup \@sanitize@url \@doi}%
  \providecommand \@doi [1]{\endgroup \@@startlink {\doibase
  #1}doi:\discretionary {}{}{}#1\@@endlink }%
}{%
  \providecommand \doi  [0]{doi:\discretionary{}{}{}\begingroup
  \urlstyle{rm}\Url }%
}%
\providecommand \doibase [0]{http://dx.doi.org/}%
\providecommand \Doi [0]{\begingroup \@sanitize@url \@Doi }%
\providecommand \@Doi  [1]{\endgroup\@@startlink{\doibase#1}\@@Doi}%
\providecommand \@@Doi [1]{#1\@@endlink}%
\providecommand \selectlanguage [0]{\@gobble}%
\providecommand \bibinfo  [0]{\@secondoftwo}%
\providecommand \bibfield  [0]{\@secondoftwo}%
\providecommand \translation [1]{[#1]}%
\providecommand \BibitemOpen [0]{}%
\providecommand \bibitemStop [0]{}%
\providecommand \bibitemNoStop [0]{.\EOS\space}%
\providecommand \EOS [0]{\spacefactor3000\relax}%
\providecommand \BibitemShut  [1]{\csname bibitem#1\endcsname}%
\bibitem [{\citenamefont {Adler}\ \emph {et~al.}(2003)\citenamefont {Adler}
  \emph {et~al.}}]{Adler:2003qi}%
  \BibitemOpen
  \bibfield  {author} {\bibinfo {author} {\bibfnamefont {S.~S.}\ \bibnamefont
  {Adler}} \emph {et~al.} (\bibinfo {collaboration} {PHENIX}),\ }\Doi
  {10.1103/PhysRevLett.91.072301} {\bibfield  {journal} {\bibinfo  {journal}
  {Phys. Rev. Lett.},\ }\textbf {\bibinfo {volume} {91}},\ \bibinfo {pages}
  {072301} (\bibinfo {year} {2003})},\ \Eprint
  {http://arxiv.org/abs/nucl-ex/0304022} {arXiv:nucl-ex/0304022} \BibitemShut
  {NoStop}%
\bibitem [{\citenamefont {Adare}\ \emph {et~al.}(2008)\citenamefont {Adare}
  \emph {et~al.}}]{Adare:2008cx}%
  \BibitemOpen
  \bibfield  {author} {\bibinfo {author} {\bibfnamefont {A.}~\bibnamefont
  {Adare}} \emph {et~al.} (\bibinfo {collaboration} {PHENIX}),\ }\Doi
  {10.1103/PhysRevLett.101.162301} {\bibfield  {journal} {\bibinfo  {journal}
  {Phys. Rev. Lett.},\ }\textbf {\bibinfo {volume} {101}},\ \bibinfo {pages}
  {162301} (\bibinfo {year} {2008})},\ \Eprint {http://arxiv.org/abs/0801.4555}
  {arXiv:0801.4555 [nucl-ex]} \BibitemShut {NoStop}%
\bibitem [{\citenamefont {Schenke}\ \emph
  {et~al.}(2010){\natexlab{a}}\citenamefont {Schenke}, \citenamefont {Jeon},\
  and\ \citenamefont {Gale}}]{Schenke:2010nt}%
  \BibitemOpen
  \bibfield  {author} {\bibinfo {author} {\bibfnamefont {B.}~\bibnamefont
  {Schenke}}, \bibinfo {author} {\bibfnamefont {S.}~\bibnamefont {Jeon}}, \
  and\ \bibinfo {author} {\bibfnamefont {C.}~\bibnamefont {Gale}},\ }\href@noop
  {} { (\bibinfo {year} {2010}{\natexlab{a}})},\ \Eprint
  {http://arxiv.org/abs/1004.1408} {arXiv:1004.1408 [hep-ph]} \BibitemShut
  {NoStop}%
\bibitem [{\citenamefont {Schenke}\ \emph
  {et~al.}(2009){\natexlab{a}}\citenamefont {Schenke}, \citenamefont {Gale},\
  and\ \citenamefont {Jeon}}]{Schenke:2009gb}%
  \BibitemOpen
  \bibfield  {author} {\bibinfo {author} {\bibfnamefont {B.}~\bibnamefont
  {Schenke}}, \bibinfo {author} {\bibfnamefont {C.}~\bibnamefont {Gale}}, \
  and\ \bibinfo {author} {\bibfnamefont {S.}~\bibnamefont {Jeon}},\ }\Doi
  {10.1103/PhysRevC.80.054913} {\bibfield  {journal} {\bibinfo  {journal}
  {Phys. Rev.},\ }\textbf {\bibinfo {volume} {C80}},\ \bibinfo {pages} {054913}
  (\bibinfo {year} {2009}{\natexlab{a}})},\ \Eprint
  {http://arxiv.org/abs/0909.2037} {arXiv:0909.2037 [hep-ph]} \BibitemShut
  {NoStop}%
\bibitem [{\citenamefont {Arnold}\ \emph {et~al.}(2001)\citenamefont {Arnold},
  \citenamefont {Moore},\ and\ \citenamefont {Yaffe}}]{Arnold:2001ms}%
  \BibitemOpen
  \bibfield  {author} {\bibinfo {author} {\bibfnamefont {P.}~\bibnamefont
  {Arnold}}, \bibinfo {author} {\bibfnamefont {G.~D.}\ \bibnamefont {Moore}}, \
  and\ \bibinfo {author} {\bibfnamefont {L.~G.}\ \bibnamefont {Yaffe}},\
  }\href@noop {} {\bibfield  {journal} {\bibinfo  {journal} {JHEP},\ }\textbf
  {\bibinfo {volume} {12}},\ \bibinfo {pages} {009} (\bibinfo {year}
  {2001})}\BibitemShut {NoStop}%
\bibitem [{\citenamefont {Arnold}\ \emph {et~al.}(2002)\citenamefont {Arnold},
  \citenamefont {Moore},\ and\ \citenamefont {Yaffe}}]{Arnold:2002ja}%
  \BibitemOpen
  \bibfield  {author} {\bibinfo {author} {\bibfnamefont {P.}~\bibnamefont
  {Arnold}}, \bibinfo {author} {\bibfnamefont {G.~D.}\ \bibnamefont {Moore}}, \
  and\ \bibinfo {author} {\bibfnamefont {L.~G.}\ \bibnamefont {Yaffe}},\
  }\href@noop {} {\bibfield  {journal} {\bibinfo  {journal} {JHEP},\ }\textbf
  {\bibinfo {volume} {06}},\ \bibinfo {pages} {030} (\bibinfo {year}
  {2002})}\BibitemShut {NoStop}%
\bibitem [{\citenamefont {Schenke}\ \emph
  {et~al.}(2009){\natexlab{b}}\citenamefont {Schenke}, \citenamefont {Gale},\
  and\ \citenamefont {Qin}}]{Schenke:2009ik}%
  \BibitemOpen
  \bibfield  {author} {\bibinfo {author} {\bibfnamefont {B.}~\bibnamefont
  {Schenke}}, \bibinfo {author} {\bibfnamefont {C.}~\bibnamefont {Gale}}, \
  and\ \bibinfo {author} {\bibfnamefont {G.-Y.}\ \bibnamefont {Qin}},\
  }\href@noop {} {\bibfield  {journal} {\bibinfo  {journal} {Phys. Rev.},\
  }\textbf {\bibinfo {volume} {C79}},\ \bibinfo {pages} {054908} (\bibinfo
  {year} {2009}{\natexlab{b}})}\BibitemShut {NoStop}%
\bibitem [{\citenamefont {Kurganov}\ and\ \citenamefont
  {Tadmor}(2000)}]{Kurganov:2000}%
  \BibitemOpen
  \bibfield  {author} {\bibinfo {author} {\bibfnamefont {A.}~\bibnamefont
  {Kurganov}}\ and\ \bibinfo {author} {\bibfnamefont {E.}~\bibnamefont
  {Tadmor}},\ }\Doi {doi:10.1006/jcph.2000.6459} {\bibfield  {journal}
  {\bibinfo  {journal} {Journal of Computational Physics},\ }\textbf {\bibinfo
  {volume} {160}},\ \bibinfo {pages} {214} (\bibinfo {year}
  {2000})}\BibitemShut {NoStop}%
\bibitem [{\citenamefont {Huovinen}\ and\ \citenamefont
  {Petreczky}(2009)}]{Huovinen:2009yb}%
  \BibitemOpen
  \bibfield  {author} {\bibinfo {author} {\bibfnamefont {P.}~\bibnamefont
  {Huovinen}}\ and\ \bibinfo {author} {\bibfnamefont {P.}~\bibnamefont
  {Petreczky}},\ }\href@noop {} { (\bibinfo {year} {2009})},\ \Eprint
  {http://arxiv.org/abs/0912.2541} {arXiv:0912.2541 [hep-ph]} \BibitemShut
  {NoStop}%
\bibitem [{\citenamefont {Miller}\ \emph {et~al.}(2007)\citenamefont {Miller},
  \citenamefont {Reygers}, \citenamefont {Sanders},\ and\ \citenamefont
  {Steinberg}}]{Miller:2007ri}%
  \BibitemOpen
  \bibfield  {author} {\bibinfo {author} {\bibfnamefont {M.~L.}\ \bibnamefont
  {Miller}}, \bibinfo {author} {\bibfnamefont {K.}~\bibnamefont {Reygers}},
  \bibinfo {author} {\bibfnamefont {S.~J.}\ \bibnamefont {Sanders}}, \ and\
  \bibinfo {author} {\bibfnamefont {P.}~\bibnamefont {Steinberg}},\ }\Doi
  {10.1146/annurev.nucl.57.090506.123020} {\bibfield  {journal} {\bibinfo
  {journal} {Ann. Rev. Nucl. Part. Sci.},\ }\textbf {\bibinfo {volume} {57}},\
  \bibinfo {pages} {205} (\bibinfo {year} {2007})},\ \Eprint
  {http://arxiv.org/abs/nucl-ex/0701025} {arXiv:nucl-ex/0701025} \BibitemShut
  {NoStop}%
\bibitem [{\citenamefont {Ishii}\ and\ \citenamefont
  {Muroya}(1992)}]{Ishii:1992xi}%
  \BibitemOpen
  \bibfield  {author} {\bibinfo {author} {\bibfnamefont {T.}~\bibnamefont
  {Ishii}}\ and\ \bibinfo {author} {\bibfnamefont {S.}~\bibnamefont {Muroya}},\
  }\Doi {10.1103/PhysRevD.46.5156} {\bibfield  {journal} {\bibinfo  {journal}
  {Phys. Rev.},\ }\textbf {\bibinfo {volume} {D46}},\ \bibinfo {pages} {5156}
  (\bibinfo {year} {1992})}\BibitemShut {NoStop}%
\bibitem [{\citenamefont {Morita}\ \emph {et~al.}(2000)\citenamefont {Morita},
  \citenamefont {Muroya}, \citenamefont {Nakamura},\ and\ \citenamefont
  {Nonaka}}]{Morita:1999vj}%
  \BibitemOpen
  \bibfield  {author} {\bibinfo {author} {\bibfnamefont {K.}~\bibnamefont
  {Morita}}, \bibinfo {author} {\bibfnamefont {S.}~\bibnamefont {Muroya}},
  \bibinfo {author} {\bibfnamefont {H.}~\bibnamefont {Nakamura}}, \ and\
  \bibinfo {author} {\bibfnamefont {C.}~\bibnamefont {Nonaka}},\ }\Doi
  {10.1103/PhysRevC.61.034904} {\bibfield  {journal} {\bibinfo  {journal}
  {Phys. Rev.},\ }\textbf {\bibinfo {volume} {C61}},\ \bibinfo {pages} {034904}
  (\bibinfo {year} {2000})},\ \Eprint {http://arxiv.org/abs/nucl-th/9906037}
  {arXiv:nucl-th/9906037} \BibitemShut {NoStop}%
\bibitem [{\citenamefont {Hirano}\ \emph {et~al.}(2002)\citenamefont {Hirano},
  \citenamefont {Morita}, \citenamefont {Muroya},\ and\ \citenamefont
  {Nonaka}}]{Hirano:2001yi}%
  \BibitemOpen
  \bibfield  {author} {\bibinfo {author} {\bibfnamefont {T.}~\bibnamefont
  {Hirano}}, \bibinfo {author} {\bibfnamefont {K.}~\bibnamefont {Morita}},
  \bibinfo {author} {\bibfnamefont {S.}~\bibnamefont {Muroya}}, \ and\ \bibinfo
  {author} {\bibfnamefont {C.}~\bibnamefont {Nonaka}},\ }\Doi
  {10.1103/PhysRevC.65.061902} {\bibfield  {journal} {\bibinfo  {journal}
  {Phys. Rev.},\ }\textbf {\bibinfo {volume} {C65}},\ \bibinfo {pages} {061902}
  (\bibinfo {year} {2002})},\ \Eprint {http://arxiv.org/abs/nucl-th/0110009}
  {arXiv:nucl-th/0110009} \BibitemShut {NoStop}%
\bibitem [{\citenamefont {Hirano}(2002)}]{Hirano:2001eu}%
  \BibitemOpen
  \bibfield  {author} {\bibinfo {author} {\bibfnamefont {T.}~\bibnamefont
  {Hirano}},\ }\Doi {10.1103/PhysRevC.65.011901} {\bibfield  {journal}
  {\bibinfo  {journal} {Phys. Rev.},\ }\textbf {\bibinfo {volume} {C65}},\
  \bibinfo {pages} {011901} (\bibinfo {year} {2002})},\ \Eprint
  {http://arxiv.org/abs/nucl-th/0108004} {arXiv:nucl-th/0108004} \BibitemShut
  {NoStop}%
\bibitem [{\citenamefont {Hirano}\ and\ \citenamefont
  {Tsuda}(2002)}]{Hirano:2002ds}%
  \BibitemOpen
  \bibfield  {author} {\bibinfo {author} {\bibfnamefont {T.}~\bibnamefont
  {Hirano}}\ and\ \bibinfo {author} {\bibfnamefont {K.}~\bibnamefont {Tsuda}},\
  }\Doi {10.1103/PhysRevC.66.054905} {\bibfield  {journal} {\bibinfo  {journal}
  {Phys. Rev.},\ }\textbf {\bibinfo {volume} {C66}},\ \bibinfo {pages} {054905}
  (\bibinfo {year} {2002})},\ \Eprint {http://arxiv.org/abs/nucl-th/0205043}
  {arXiv:nucl-th/0205043} \BibitemShut {NoStop}%
\bibitem [{\citenamefont {Morita}\ \emph {et~al.}(2002)\citenamefont {Morita},
  \citenamefont {Muroya}, \citenamefont {Nonaka},\ and\ \citenamefont
  {Hirano}}]{Morita:2002av}%
  \BibitemOpen
  \bibfield  {author} {\bibinfo {author} {\bibfnamefont {K.}~\bibnamefont
  {Morita}}, \bibinfo {author} {\bibfnamefont {S.}~\bibnamefont {Muroya}},
  \bibinfo {author} {\bibfnamefont {C.}~\bibnamefont {Nonaka}}, \ and\ \bibinfo
  {author} {\bibfnamefont {T.}~\bibnamefont {Hirano}},\ }\Doi
  {10.1103/PhysRevC.66.054904} {\bibfield  {journal} {\bibinfo  {journal}
  {Phys. Rev.},\ }\textbf {\bibinfo {volume} {C66}},\ \bibinfo {pages} {054904}
  (\bibinfo {year} {2002})},\ \Eprint {http://arxiv.org/abs/nucl-th/0205040}
  {arXiv:nucl-th/0205040} \BibitemShut {NoStop}%
\bibitem [{\citenamefont {Nonaka}\ and\ \citenamefont
  {Bass}(2007)}]{Nonaka:2006yn}%
  \BibitemOpen
  \bibfield  {author} {\bibinfo {author} {\bibfnamefont {C.}~\bibnamefont
  {Nonaka}}\ and\ \bibinfo {author} {\bibfnamefont {S.~A.}\ \bibnamefont
  {Bass}},\ }\Doi {10.1103/PhysRevC.75.014902} {\bibfield  {journal} {\bibinfo
  {journal} {Phys. Rev.},\ }\textbf {\bibinfo {volume} {C75}},\ \bibinfo
  {pages} {014902} (\bibinfo {year} {2007})}\BibitemShut {NoStop}%
\bibitem [{\citenamefont {Niemi}\ \emph {et~al.}(2009)\citenamefont {Niemi},
  \citenamefont {Eskola},\ and\ \citenamefont {Ruuskanen}}]{Niemi:2008ta}%
  \BibitemOpen
  \bibfield  {author} {\bibinfo {author} {\bibfnamefont {H.}~\bibnamefont
  {Niemi}}, \bibinfo {author} {\bibfnamefont {K.~J.}\ \bibnamefont {Eskola}}, \
  and\ \bibinfo {author} {\bibfnamefont {P.~V.}\ \bibnamefont {Ruuskanen}},\
  }\Doi {10.1103/PhysRevC.79.024903} {\bibfield  {journal} {\bibinfo  {journal}
  {Phys. Rev.},\ }\textbf {\bibinfo {volume} {C79}},\ \bibinfo {pages} {024903}
  (\bibinfo {year} {2009})},\ \Eprint {http://arxiv.org/abs/0806.1116}
  {arXiv:0806.1116 [hep-ph]} \BibitemShut {NoStop}%
\bibitem [{\citenamefont {Cooper}\ and\ \citenamefont
  {Frye}(1974)}]{Cooper:1974mv}%
  \BibitemOpen
  \bibfield  {author} {\bibinfo {author} {\bibfnamefont {F.}~\bibnamefont
  {Cooper}}\ and\ \bibinfo {author} {\bibfnamefont {G.}~\bibnamefont {Frye}},\
  }\Doi {10.1103/PhysRevD.10.186} {\bibfield  {journal} {\bibinfo  {journal}
  {Phys. Rev.},\ }\textbf {\bibinfo {volume} {D10}},\ \bibinfo {pages} {186}
  (\bibinfo {year} {1974})}\BibitemShut {NoStop}%
\bibitem [{\citenamefont {Jeon}\ and\ \citenamefont
  {Moore}(2005)}]{Jeon:2003gi}%
  \BibitemOpen
  \bibfield  {author} {\bibinfo {author} {\bibfnamefont {S.}~\bibnamefont
  {Jeon}}\ and\ \bibinfo {author} {\bibfnamefont {G.~D.}\ \bibnamefont
  {Moore}},\ }\Doi {10.1103/PhysRevC.71.034901} {\bibfield  {journal} {\bibinfo
   {journal} {Phys. Rev.},\ }\textbf {\bibinfo {volume} {C71}},\ \bibinfo
  {pages} {034901} (\bibinfo {year} {2005})}\BibitemShut {NoStop}%
\bibitem [{\citenamefont {Fries}\ \emph {et~al.}(2003)\citenamefont {Fries},
  \citenamefont {Muller},\ and\ \citenamefont {Srivastava}}]{Fries:2002kt}%
  \BibitemOpen
  \bibfield  {author} {\bibinfo {author} {\bibfnamefont {R.~J.}\ \bibnamefont
  {Fries}}, \bibinfo {author} {\bibfnamefont {B.}~\bibnamefont {Muller}}, \
  and\ \bibinfo {author} {\bibfnamefont {D.~K.}\ \bibnamefont {Srivastava}},\
  }\Doi {10.1103/PhysRevLett.90.132301} {\bibfield  {journal} {\bibinfo
  {journal} {Phys. Rev. Lett.},\ }\textbf {\bibinfo {volume} {90}},\ \bibinfo
  {pages} {132301} (\bibinfo {year} {2003})}\BibitemShut {NoStop}%
\bibitem [{\citenamefont {Turbide}\ \emph {et~al.}(2008)\citenamefont
  {Turbide}, \citenamefont {Gale}, \citenamefont {Frodermann},\ and\
  \citenamefont {Heinz}}]{Turbide:2007mi}%
  \BibitemOpen
  \bibfield  {author} {\bibinfo {author} {\bibfnamefont {S.}~\bibnamefont
  {Turbide}}, \bibinfo {author} {\bibfnamefont {C.}~\bibnamefont {Gale}},
  \bibinfo {author} {\bibfnamefont {E.}~\bibnamefont {Frodermann}}, \ and\
  \bibinfo {author} {\bibfnamefont {U.}~\bibnamefont {Heinz}},\ }\Doi
  {10.1103/PhysRevC.77.024909} {\bibfield  {journal} {\bibinfo  {journal}
  {Phys. Rev.},\ }\textbf {\bibinfo {volume} {C77}},\ \bibinfo {pages} {024909}
  (\bibinfo {year} {2008})}\BibitemShut {NoStop}%
\bibitem [{\citenamefont {Whalley}\ \emph {et~al.}(2005)\citenamefont
  {Whalley}, \citenamefont {Bourilkov},\ and\ \citenamefont
  {Group}}]{Whalley:2005nh}%
  \BibitemOpen
  \bibfield  {author} {\bibinfo {author} {\bibfnamefont {M.~R.}\ \bibnamefont
  {Whalley}}, \bibinfo {author} {\bibfnamefont {D.}~\bibnamefont {Bourilkov}},
  \ and\ \bibinfo {author} {\bibfnamefont {R.~C.}\ \bibnamefont {Group}},\
  }\href@noop {} { (\bibinfo {year} {2005})},\ \Eprint
  {http://arxiv.org/abs/hep-ph/0508110} {arXiv:hep-ph/0508110} \BibitemShut
  {NoStop}%
\bibitem [{\citenamefont {Eskola}\ \emph {et~al.}(1999)\citenamefont {Eskola},
  \citenamefont {Kolhinen},\ and\ \citenamefont {Salgado}}]{Eskola:1998df}%
  \BibitemOpen
  \bibfield  {author} {\bibinfo {author} {\bibfnamefont {K.~J.}\ \bibnamefont
  {Eskola}}, \bibinfo {author} {\bibfnamefont {V.~J.}\ \bibnamefont
  {Kolhinen}}, \ and\ \bibinfo {author} {\bibfnamefont {C.~A.}\ \bibnamefont
  {Salgado}},\ }\Doi {10.1007/s100520050513} {\bibfield  {journal} {\bibinfo
  {journal} {Eur. Phys. J.},\ }\textbf {\bibinfo {volume} {C9}},\ \bibinfo
  {pages} {61} (\bibinfo {year} {1999})}\BibitemShut {NoStop}%
\bibitem [{\citenamefont {Andersson}\ \emph {et~al.}(1983)\citenamefont
  {Andersson}, \citenamefont {Gustafson}, \citenamefont {Ingelman},\ and\
  \citenamefont {Sjostrand}}]{Andersson:1983ia}%
  \BibitemOpen
  \bibfield  {author} {\bibinfo {author} {\bibfnamefont {B.}~\bibnamefont
  {Andersson}}, \bibinfo {author} {\bibfnamefont {G.}~\bibnamefont
  {Gustafson}}, \bibinfo {author} {\bibfnamefont {G.}~\bibnamefont {Ingelman}},
  \ and\ \bibinfo {author} {\bibfnamefont {T.}~\bibnamefont {Sjostrand}},\
  }\Doi {10.1016/0370-1573(83)90080-7} {\bibfield  {journal} {\bibinfo
  {journal} {Phys. Rept.},\ }\textbf {\bibinfo {volume} {97}},\ \bibinfo
  {pages} {31} (\bibinfo {year} {1983})}\BibitemShut {NoStop}%
\bibitem [{\citenamefont {Sjostrand}(1984)}]{Sjostrand:1984ic}%
  \BibitemOpen
  \bibfield  {author} {\bibinfo {author} {\bibfnamefont {T.}~\bibnamefont
  {Sjostrand}},\ }\Doi {10.1016/0550-3213(84)90607-2} {\bibfield  {journal}
  {\bibinfo  {journal} {Nucl. Phys.},\ }\textbf {\bibinfo {volume} {B248}},\
  \bibinfo {pages} {469} (\bibinfo {year} {1984})}\BibitemShut {NoStop}%
\bibitem [{\citenamefont {Chujo}(2007)}]{Chujo:2007hi}%
  \BibitemOpen
  \bibfield  {author} {\bibinfo {author} {\bibfnamefont {T.}~\bibnamefont
  {Chujo}} (\bibinfo {collaboration} {PHENIX}),\ }\Doi
  {10.1088/0954-3899/34/8/S120} {\bibfield  {journal} {\bibinfo  {journal} {J.
  Phys.},\ }\textbf {\bibinfo {volume} {G34}},\ \bibinfo {pages} {S893}
  (\bibinfo {year} {2007})},\ \Eprint {http://arxiv.org/abs/nucl-ex/0703014}
  {arXiv:nucl-ex/0703014} \BibitemShut {NoStop}%
\bibitem [{\citenamefont {Abelev}\ \emph {et~al.}(2010)\citenamefont {Abelev}
  \emph {et~al.}}]{Abelev:2010tr}%
  \BibitemOpen
  \bibfield  {author} {\bibinfo {author} {\bibfnamefont {B.~I.}\ \bibnamefont
  {Abelev}} \emph {et~al.} (\bibinfo {collaboration} {The STAR}),\ }\Doi
  {10.1103/PhysRevC.81.044902} {\bibfield  {journal} {\bibinfo  {journal}
  {Phys. Rev.},\ }\textbf {\bibinfo {volume} {C81}},\ \bibinfo {pages} {044902}
  (\bibinfo {year} {2010})},\ \Eprint {http://arxiv.org/abs/1001.5052}
  {arXiv:1001.5052 [nucl-ex]} \BibitemShut {NoStop}%
\bibitem [{\citenamefont {Schenke}\ \emph
  {et~al.}(2010){\natexlab{b}}\citenamefont {Schenke}, \citenamefont {Jeon},\
  and\ \citenamefont {Gale}}]{Schenke:2010rr}%
  \BibitemOpen
  \bibfield  {author} {\bibinfo {author} {\bibfnamefont {B.}~\bibnamefont
  {Schenke}}, \bibinfo {author} {\bibfnamefont {S.}~\bibnamefont {Jeon}}, \
  and\ \bibinfo {author} {\bibfnamefont {C.}~\bibnamefont {Gale}},\ }\href@noop
  {} { (\bibinfo {year} {2010}{\natexlab{b}})},\ \Eprint
  {http://arxiv.org/abs/1009.3244} {arXiv:1009.3244 [hep-ph]} \BibitemShut
  {NoStop}%
\bibitem [{\citenamefont {Veres}\ \emph {et~al.}(2008)\citenamefont {Veres}
  \emph {et~al.}}]{Veres:2008nq}%
  \BibitemOpen
  \bibfield  {author} {\bibinfo {author} {\bibfnamefont {G.~I.}\ \bibnamefont
  {Veres}} \emph {et~al.} (\bibinfo {collaboration} {PHOBOS Collaboration}),\
  }\href@noop {} { (\bibinfo {year} {2008})},\ \Eprint
  {http://arxiv.org/abs/0806.2803} {arXiv:0806.2803 [nucl-ex]} \BibitemShut
  {NoStop}%
\bibitem [{\citenamefont {Abelev}\ \emph {et~al.}(2009)\citenamefont {Abelev}
  \emph {et~al.}}]{Abelev:2009ue}%
  \BibitemOpen
  \bibfield  {author} {\bibinfo {author} {\bibfnamefont {B.~I.}\ \bibnamefont
  {Abelev}} \emph {et~al.} (\bibinfo {collaboration} {STAR}),\ }\href@noop {} {
  (\bibinfo {year} {2009})},\ \Eprint {http://arxiv.org/abs/0911.3130}
  {arXiv:0911.3130 [nucl-ex]} \BibitemShut {NoStop}%
\end{thebibliography}%

\end{document}